\documentclass[%
 reprint,showkeys,showpacs,amsmath,amssymb,aps,prd]{revtex4-1}
\usepackage{graphicx}
\usepackage[caption=false]{subfig}
\usepackage{color}
\usepackage[utf8]{inputenc}
\usepackage{amssymb}
\usepackage[pdftex]{hyperref}
\begin{document}
\title{Cylindrical spacetimes due to radial magnetic fields}
\author{Ji\v{r}\'{i} \surname{Vesel\'{y}}}%
\email{jiri.vesely@utf.mff.cuni.cz}
\author{Martin \surname{\v{Z}ofka}}
\email{zofka@mbox.troja.mff.cuni.cz}
\affiliation{Institute of Theoretical Physics, Faculty of Mathematics and Physics, Charles University, Czech Republic}
\vspace{10pt}
\begin{abstract}
We continue our previous study of cylindrically symmetric, static electrovacuum spacetimes generated by a magnetic field, involving optionally the cosmological constant, and investigate several classes of exact solutions. These spacetimes are due to magnetic fields that are perpendicular to the axis of symmetry.
\end{abstract}
\maketitle
\section{Introduction}\label{Introduction}
Electromagnetic fields play an important role not only in astrophysics---they are prominent in neutron stars and affect entire galaxies---and even in cosmology---where the primordial fields give rise to the observed intergalactic magnetic fields---but also in mathematical general relativity. Indeed, one of the first exact non-vacuum solutions of Einstein equations was the Reissner-Nordstr\"{o}m spacetime featuring a black hole endowed with both mass and electric charge. Another of the many interesting exact solutions of Einstein-Maxwell equations is the Bonnor-Melvin universe \cite{Bonnor,Melvin} that, unlike the Reissner-Nordstr\"{o}m solution, is cylindrically symmetric. It describes the gravitational field generated by an axial magnetic field permeating the whole spacetime and it can be thought of as due to azimuthal current on the surface of a coaxial cylinder enveloping part of the spacetime.

Although cylindrical symmetry requires objects of infinite extension and thus is not of direct interest in astrophysics, it still plays an important role in the collapse of rotating objects which can produce spindle-like structures (see, e.g., \cite{Yoo+Harada+Okawa,East}) approximated in their vicinity by cylindrical symmetry. On a bigger scale, one has the cosmic strings, which might have arisen due to phase transitions in the early universe and which are locally cylindrically symmetric, see \cite{Copeland+Kibble}. Another reason for imposing cylindrical symmetry is that it is generally difficult to find exact solutions of Einstein equations---enforcing any symmetry on the gravitational field reduces them to a more tractable system and any exact solution provides us with an insight into what might be relevant in a more realistic situation. In this paper, we thus study several exact solutions of Einstein-Maxwell equations closely related to the Bonnor-Melvin solution through their symmetry and the presence of the electromagnetic field. One cannot simply apply the hoop conjecture \cite{Thorne,Hod} to an infinite object and, indeed, we often end up with naked singularities we also discuss solutions featuring a singularity covered by a horizon. Our solutions are not asymptotically flat---they cannot be since they are translationally invariant along their axis. We thus look at their behavior along the radial cylindrical coordinate instead. One is of a finite proper extent and the rest is asymptotically either Minkowski, de Sitter or anti de Sitter according to the value of the cosmological constant, which we generally include in our calculations.

The paper consists of several sections: we first set the stage by defining our coordinate system and the general form of the metric with the corresponding Einstein-Maxwell equations. At this point we still assume a magnetic field aligned with the symmetry axis. In Section \ref{The_other_non-cosmological_solution} we examine the case with a vanishing cosmological constant, concluding that, apart from the Bonnor-Melvin case, it leads us to a spacetime with a magnetic field perpendicular to the axis and we discuss its properties. We thus continue with Section \ref{Self-gravitating_radial_magnetic_field} giving the set of Einstein-Maxwell equations describing spacetimes with a radial magnetic field. Sections \ref{Homogeneous_solution_with_Lambda}, \ref{Inhomogeneous_solution_with_Lambda}, and \ref{Inhomogeneous_solution_without_Lambda} then study solutions with a homogeneous magnetic field and an inhomogeneous magnetic field with and without the cosmological constant, respectively. We conclude with some open questions and summarize our results.

\section{The setting}\label{The_setting}
Following our previous paper \cite{Zofka}, where we generalized the Bonnor-Melvin solution to yield a homogeneous magnetic field and include the cosmological constant, and the solutions allowing a variation of the magnetic field \cite{Astorino,Lim} which we found independently in \cite{Vesely+Zofka}, we write the metric of a general static, cylindrically symmetric spacetime as
\begin{equation}\label{general line element}
\mathrm{d}s^2 = -\exp A(r) \; \mathrm{d}t^2 + \mathrm{d}r^2 + \exp B(r) \; \mathrm{d}z^2 + \exp C(r) \;\mathrm{d}\varphi^2,
\end{equation}
where $r \in \! I\hspace{-0.13cm}R^+$ is the proper radial distance, $t,z \in \! I\hspace{-0.13cm}R$ are temporal and azimuthal coordinates, and $\varphi \in [0,2\pi)$ measures the angle around the axis of symmetry but we give it the dimension of length for the metric to be consistent. We are looking for a self-consistent solution generated by a magnetic field aligned with the axis of symmetry
\begin{equation}\label{Maxwell original}
    F = H(r) \: \mbox{d}r \wedge \mbox{d}\varphi,
\end{equation}
yielding
\begin{equation}\label{Maxwell invariant}
    F_{\mu\nu}F^{\mu\nu} = 2H^2\mbox{e}^{-C} =: 2f^2,
\end{equation}
where we defined a new quantity, $f(r)$. Einstein-Maxwell equations can then be reduced to the form
\begin{eqnarray}
  &f'''f'& - 2(f'')^2 + f''\left(6f(\Lambda-f^2) + \frac{(f')^2}{f}\right) \nonumber\\
  && + (f')^2(11f^2 - 9\Lambda) - 4f^2(\Lambda-f^2)^2=0, \label{full equation for f}
\end{eqnarray}
with primes denoting derivatives with respect to the radial coordinate, $r$, while the metric functions are obtained by integration and read
\begin{eqnarray}
  A' &=& -\frac{f'}{f} \pm \sqrt{4\frac{f''}{f}-7\left(\frac{f'}{f}\right)^2-4(\Lambda-f^2)},\label{differential equation for A}\\
  B' &=& -\frac{f'}{f} \mp \sqrt{4\frac{f''}{f}-7\left(\frac{f'}{f}\right)^2-4(\Lambda-f^2)},\label{differential equation for B}\\
C' &=& -2 \frac{f''}{f'} + 4 \frac{f'}{f} + 4\frac{f}{f'}(\Lambda-f^2).\label{differential equation for C}
\end{eqnarray}
Since (\ref{full equation for f}) does not involve the independent variable, we can reduce the order of the equation to obtain
\begin{eqnarray}
    &w \ddot{w}& - \dot{w}^2 + \left[\frac{w}{f}-6 f \left(f^2-\Lambda\right)\right] \dot{w} \nonumber\\
    && - 8 \left(f^2-\Lambda\right)^2 f^2 + 2 w \left( 11 f^2 - 9 \Lambda \right) = 0, \label{full equation for w}
\end{eqnarray}
with dots denoting derivatives with respect to $f$ and $w(f) := [\mathrm{d}f(r)/\mathrm{d}r]^2 $. We now explore the solutions of the above set of equations.
\section{The other non-cosmological solution}\label{The_other_non-cosmological_solution}
Let us begin by reexamining the non-cosmological case. We set $\Lambda=0$ and solve (\ref{full equation for w}) to obtain an exact solution of the form
\begin{equation}\label{the other BM}
  w = -4 f^4 + 4\alpha f^3,
\end{equation}
which is different from the Bonnor-Melvin metric with $w = -4 f^4 + \gamma f^{7/2}$ \cite{Bonnor,Melvin}, see the discussion in \cite{Vesely+Zofka}. Since $w = (f')^2$, we can separate variables to find
\begin{equation}\label{function f}
  f = \frac{\alpha}{1+ \alpha^2 r^2}.
\end{equation}
This solves Einstein equations but there is a problem with the square root in (\ref{differential   equation for A}) and (\ref{differential   equation for B}) since they are taken from a negative number as
\begin{equation}
  4\frac{f''}{f}-7\left(\frac{f'}{f}\right)^2 + 4f^2 = -\frac{4 \alpha ^2}{1 + \alpha ^2 r^2} <0,
\end{equation}
so that we end up with a complex metric. To work around this, we take $\alpha$ to be imaginary and factor the imaginary unit out, $\alpha \rightarrow i \alpha$. This yields a real metric
\label{real metric}
\begin{eqnarray}\label{general line element original radial coordinate}
    ds^2 &=& (1 - \alpha^2 r^2) \exp (2 \arcsin \alpha r) dt^2 + dr^2 \nonumber\\
    - (1 - \alpha^2 r^2) &\exp& (-2 \arcsin \alpha r) \; dz^2 + \frac{d\varphi^2}{1 - \alpha^2 r^2},
\end{eqnarray}
but, instead, we get an imaginary $f$ through (\ref{function f})---it corresponds to the components of the Maxwell tensor with $F_{r\varphi} = f \sqrt{g_{\varphi\varphi}} = i \alpha / (1 - \alpha^2 r^2)^{3/2}$. It then suggests itself to use a Wick rotation $\varphi \rightarrow i t, t \rightarrow \varphi, z \rightarrow i z$ to return to real values and correct metric signature
\begin{eqnarray}\label{general line element original radial coordinate Wicked}
ds^2 & = & -\frac{dt^2}{1 - \alpha^2 r^2} + dr^2 + (1 - \alpha^2 r^2) \times \nonumber\\
 \left[ \right. \exp (-2 &\arcsin& \alpha r) \; dz^2 + \exp (2 \arcsin \alpha r)\; d\varphi^2 \left. \right].
\end{eqnarray}
The Maxwell field becomes real and purely electric with $F_{tr} = \alpha / (1 - \alpha^2 r^2)^{3/2}$ and $A_t = -\alpha r/\sqrt{1 - \alpha^2 r^2}$, which is no longer axial and becomes radial instead. Using the dual rotation, we finally obtain a purely magnetic field
\begin{equation}\label{Maxwell magnetic}
    F_{z \varphi} = \alpha, \;\;\; A_\varphi = \alpha z.
\end{equation}

\begin{figure}[ht]
\centering
\includegraphics[width = 0.4\textwidth]{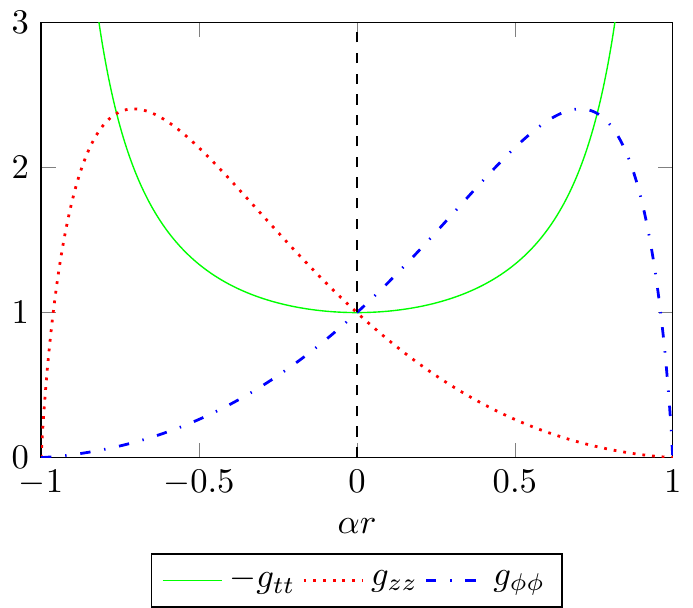}
\caption{Metric components (\ref{general line element original radial coordinate Wicked}) as functions of the radial coordinate.\label{plot: Asymmetric case original radial coord}}
\end{figure}
\begin{figure}[h]
\centering
\includegraphics[width = 0.25\textwidth]{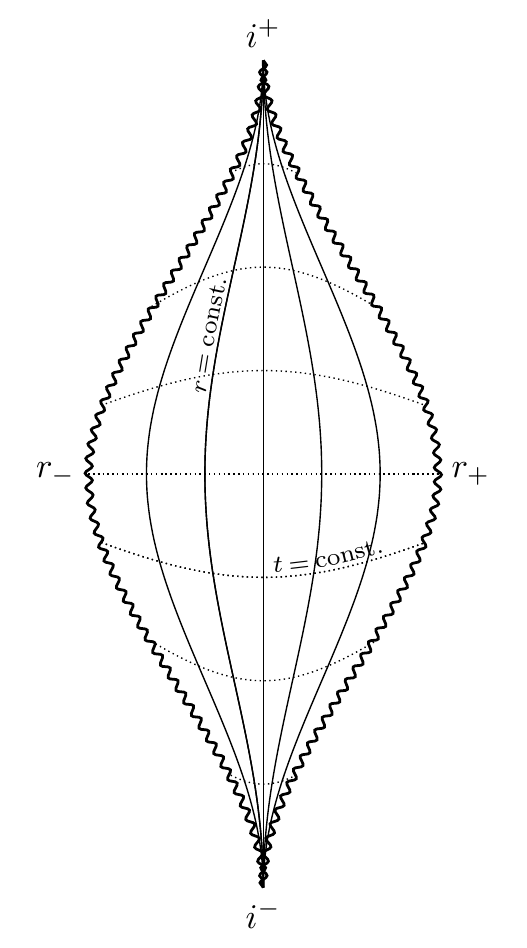}
\caption{Conformal diagram of the spacetime due to the metric (\ref{general line element original radial coordinate Wicked}). Each point represents a cylindrical surface. The singularities located at $\alpha r_\pm = \pm 1$ and indicated by wavy lines are in fact point-like as their proper length along the z-axis vanishes. Lines of constant $r$ are solid and lines of constant $t$ are dotted.\label{Penrose no Lambda}}
\end{figure}

This exact solution is a special case of (3.16) in \cite{Bronnikov+Santos+Wang}. Let us now investigate its properties in more detail. The coordinate $r$ ranges from $-1/|\alpha|$ to $1/|\alpha|$ with the Kretschmann scalar
\begin{equation}\label{Kretschmann original coords}
  K = \frac{8 \alpha^4 \left( 4 \alpha^4 r^4 + 2 \alpha^2 r^2+1 \right)}{\left( \alpha^2 r^2 - 1 \right)^4}
\end{equation}
diverging at the endpoints of the interval, which thus are physical singularities of the spacetime. Likewise, the Maxwell invariant reads
\begin{equation}
    F_{\mu \nu} F^{\mu \nu} = \frac{2 \alpha^2}{(1 - \alpha^2 r^2)^2}.
\end{equation}
It is apparent from the plot of the azimuthal metric coefficient $g_{\varphi \varphi}$, see Figure \ref{plot: Asymmetric case original radial coord}, that $\alpha r = \pm 1$ are axes where the circumference of hoops around them vanishes. A better picture is that of the globe with standard geographic coordinates---latitude ($r$) and longitude ($\varphi$). At the poles the proper length along the axis vanishes as well so that they resemble points rather than lines. In view of the fact that this is an electrovacuum solution, its source must reside within the two point singularities so that the magnetic field is analogous to that of two opposite magnetic monopoles. The corresponding Penrose diagram is presented in Figure \ref{Penrose no Lambda}.

The solution only has the 3 obvious Killing vector fields corresponding to the staticity and cylindrical symmetry. It is further invariant under the exchange $r \leftrightarrow -r, z \leftrightarrow \varphi$. Applying $r \leftrightarrow -r, \alpha \leftrightarrow -\alpha$ or exchanging $\varphi \leftrightarrow z$, the metric remains unchanged while the magnetic field changes sign. The spacetime is type I everywhere apart from $r=0$ where it is type O and $\alpha r = \pm 1/\sqrt{10}$ where it is type D. The flat space limit is achieved simply by taking $\alpha \rightarrow 0$.

Using Cartesian-like coordinates with $r^2 = x^2 + y^2$, we have
\begin{equation}\label{Maxwell in Cartesian coordinates}
  F_{zx}=-\frac{\alpha y}{r^2}=-B_{y}, \;\;\; F_{zy}=\frac{\alpha x}{r^2}=B_{x}.
\end{equation}
This implies
$$\vec{B} = \alpha \frac{\vec{e_r}}{r},$$
which shows the magnetic field is cylindrically radial.

Transforming metric (\ref{general line element original radial coordinate Wicked}) to a new radial coordinate with $1 - \alpha^2 r^2 =: \cos^2 (\alpha \rho)$ and $\alpha\rho \in [-\pi/2,\pi/2]$ (the endpoints are the two singular axes), we find
\begin{eqnarray}\label{general line element in x}
ds^2 &=& -\frac{dt^2}{\cos^2 (\alpha \rho)} + \cos^2 (\alpha \rho) d\rho^2 \nonumber\\
&&+\cos^2 (\alpha \rho) \left[ e^{-2 \alpha \rho} \; dz^2 + e^{2 \alpha \rho}\; d\varphi^2 \right],
\end{eqnarray}
with still
\begin{equation}\label{Maxwell magnetic in x}
    F_{z \varphi} = \alpha.
\end{equation}

Let us look at the motion of test particles in the gravitational and electromagnetic fields due to (\ref{general line element in x}). For uncharged particles following general geodesics, we can derive an effective potential
\begin{equation}
 V(\rho) := - \frac{1}{2g_{\rho\rho}} \left[ \delta - \frac{E^2}{g_{tt}} - \frac{Z^2}{g_{zz}} - \frac{L^2}{g_{\varphi\varphi}} \right],
\end{equation}
with $g_{\alpha\beta}$ covariant metric coefficients, $E, Z, L$ constants of motion due to the Killing vector fields $\partial/\partial t, \partial/\partial z, \partial/\partial \varphi$, and $\delta = -1, 0$ for massive particles and photons, respectively. The above potential governs geodetic motion through
\begin{equation}
 \frac{1}{2} \dot{\rho}^2 = - V(\rho) \, ,
\end{equation}
so that there are radial turning points for geodetic particles where $V(\rho)=0$. Apart from radial null geodesics, the potential diverges at $\alpha\rho = \pm \pi/2$ and, consequently, only radially moving photons can reach the singularities while all massive particles are pushed away from the singularities, which thus act repulsively, similarly to, e.g., the Kerr-Newman-(anti-)de Sitter solution \cite{Vesely+Zofka_geodesics}. Moreover, all timelike radial geodesics oscillate through $\rho=0$. For massive particles, circular orbits in the planes perpendicular to the axis can only occur for a finite range of radii, $\alpha \rho \in (0,\arctan(1/2))$. The lower endpoint of the interval yields a static massive particle (this applies even to charged particles) while the upper endpoint corresponds to circular photonic geodesics. Massive particles can also spiral along the cylinder located at $\rho=0$ (its circumference is finite). All these paths are stable under radial perturbations.

It is of interest that we lose some of the symmetry for charged particles: particles moving originally within planes perpendicular to the axes with $\dot{\varphi} \not = 0$ are pushed along the axes. Consequently, there are no circular electrogeodesics. Purely radial motion can be integrated analytically but the resulting formulas are rather unwieldy. We no longer have the integral of motion $Z$ originally due to the Killing field $\partial/\partial z$ since the electromagnetic 4-potential contains $z$ and charged particles even exhibit chaotic motion for certain initial conditions unlike in the original Bonnor-Melvin spacetime.

The gravitational and magnetic fields (\ref{general line element in x}) and (\ref{Maxwell magnetic in x}) can be thought of as due to infinitely thin cylindrical shells of charged matter aligned with the axes. We studied the case where we replace part of the spacetime (\ref{general line element original radial coordinate Wicked}) by either Minkowski or the standard Bonnor-Melvin solution. We can always find such a combination of parameters that the resulting 3D induced energy-momentum tensor can be interpreted as due to 4 counter-spiralling streams of massive and charged particles. However, we cannot cut out both axes at the same time in this way and the ensuing spacetime thus always contains a singularity.
\section{Self-gravitating radial magnetic field}\label{Self-gravitating_radial_magnetic_field}
Inspired by our previous considerations, we now return to the original system (\ref{general line element}) but endowed with a radial magnetic field
\begin{equation}\label{Maxwell original radial}
    F = H(r) \exp \frac{B(r)}{2}\: \mbox{d}z \wedge \mbox{d}\varphi,
\end{equation}
yielding
\begin{equation}\label{Maxwell invariant radial}
    F_{\mu\nu}F^{\mu\nu} = 2H^2\mbox{e}^{-C} =: 2f^2,
\end{equation}
where we again introduced $f(r)$, while we still find $\star F_{\mu \nu} F^{\mu \nu} = 0$. Einstein-Maxwell equations now read
\begin{eqnarray}
  2(B'' + C'') + \left( B' \right)^{2} + \left( C' \right) ^{2} + B'C' + 4\Lambda + 4f^2 &=& 0,\;\;\;\;\;\label{Einstein t radial}\\
  2(A'' + C'') + \left( A' \right)^{2} + \left( C' \right) ^{2} + A'C' + 4\Lambda - 4f^2 &=& 0,\label{Einstein z radial}\\
  2(A'' + B'') + \left( A' \right)^{2} + \left( B' \right) ^{2} + A'B' + 4\Lambda - 4f^2 &=& 0,\label{Einstein phi radial}\\
  A'B' + A'C' + B'C' + 4\Lambda + 4f^2 &=& 0.\label{Einstein r radial}
\end{eqnarray}
Proceding along the lines of the axial case, we differentiate (\ref{Einstein r radial}), multiply it by 2, and subtract from it $A'$.(\ref{Einstein t radial}) + $B'$.(\ref{Einstein z radial}) + $C'$.(\ref{Einstein phi radial}) to obtain
\begin{eqnarray}
    &&16ff' + 4f^2(-A+B+C)' \nonumber\\
    &&-(A+B+C)'(4\Lambda + A'B' + A'C' + B'C') =0,
\end{eqnarray}
where we substitute for the last bracket from (\ref{Einstein r radial}) to yield
\begin{equation}
    2f' + f(B+C)' = 0,
\end{equation}
which can be integrated to yield
\begin{equation}\label{Maxwell constraint radial}
    \mbox{e}^{\frac{B+C}{2}} f = const.
\end{equation}

\section{``Homogeneous'' solution with \texorpdfstring{$\Lambda<0$}{L<0}}\label{Homogeneous_solution_with_Lambda}
\begin{figure}[h]
\centering
\includegraphics[width = 0.25\textwidth]{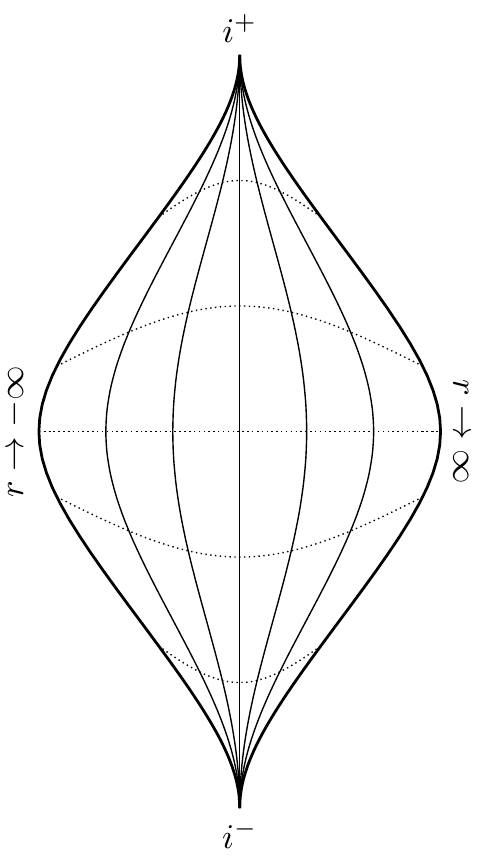}
\caption{Conformal diagram of the homogeneous spacetime (\ref{metric with deficit angle radial}). Each point represents a cylindrical surface (or a planar one, depending on the definition of the coordinate $\varphi$). Solid and dotted lines represent constant $r$ and $t$, respectively.\label{Penrose homogeneous with Lambda}}
\end{figure}

Following our previous work \cite{Zofka}, we start with the ``homogeneous'' case $f=$ const., which yields immediately $B$ and $C$ constant and
\begin{equation}\label{radial}
  f^2 = -\Lambda>0,
\end{equation}
with a negative cosmological constant. The only remaining Einstein equation reads
\begin{equation}\label{equation for A radial}
  2 A'' + A'^2 + 8\Lambda =0,
\end{equation}
so that
\begin{equation}\label{A radial}
  A = 2 \log \left[ \alpha \cosh \sqrt{-2\Lambda}(r-R) \right].
\end{equation}
Rescaling $t$ and $z$ and shifting $r$, this translates into
\begin{equation}\label{metric with deficit angle radial}
ds^2 = - \cosh^2 \left(\sqrt{-2 \Lambda} r \right) dt^2 + dr^2 + dz^2 + \sigma^2 d\varphi^2,
\end{equation}
generally with conicity due to the presence of $\sigma$ while the magnetic field reads
\begin{equation}
H(r) = \sigma \sqrt{-\Lambda}, \;\;\; F_{z \varphi} = \sigma \sqrt{-\Lambda}, \;\;\; A_\varphi =  \sigma \sqrt{-\Lambda} z.
\end{equation}
This is in fact an $AdS_2 \times I\hspace{-0.13cm}R_2$ space, or the ``exceptional electrovacuum type D Kundt metric with cosmological constant'' investigated by Pleba\'{n}ski and Hacyan \cite{Plebanski+Hacyan}, see also \cite{Griffiths+Podolsky}. It is completely analogous to the homogeneous axial solution we have discussed previously in \cite{Zofka} and it describes an electromagnetic field held together entirely by its own gravity. See, Figure \ref{Penrose homogeneous with Lambda} for the corresponding Penrose diagram.
\section{``Inhomogeneous'' symmetric solution with a non-vanishing \texorpdfstring{$\Lambda$}{L}}\label{Inhomogeneous_solution_with_Lambda}

To deal with the non-homogeneous case, we first assume a symmetric metric with $B=C$ to simplify Einstein equations (\ref{Einstein t radial})-(\ref{Einstein r radial}). In analogy with our previous work \cite{Vesely+Zofka}, we find
\begin{eqnarray}
\mathrm{d}s^2 &=& - \frac{\gamma f^{\frac{7}{2}} + 4 f^4 - \frac{4}{3}\Lambda f^2}{f^3} \mathrm{d}t^2 + \frac{\mathrm{d}f^2}{\gamma f^{\frac{7}{2}} + 4 f^4 - \frac{4}{3}\Lambda f^2} \nonumber \\
&&+ \frac{1}{f} \left(\mathrm{d}z^2 + \beta^2 \mathrm{d}\varphi^2 \right)\label{Inhomogeneous solution with Lambda}
\end{eqnarray}
while the magnetic field reads
\begin{equation}
H(f) = \beta \sqrt{f}, \;\;\; F_{z \varphi} = \beta, \;\;\; A_\varphi = \beta z.
\end{equation}
The units of the coordinates and variables appearing in (\ref{Inhomogeneous solution with Lambda}) are $[f]=m^{-1}, [t]=m^{3/2}, [z]=[\varphi]=m^{1/2}$ with $[\gamma]=m^{-1/2}, [\beta]=1, [\Lambda]=m^{-2}$. Unlike with an axial magnetic field, we now have $\mathrm{sgn}(g_{tt}) = \mathrm{sgn}(g_{ff})$ and the required $+2$ signature of the metric (\ref{Inhomogeneous solution with Lambda}) thus admits both signs of $g_{ff}$, reminiscent of the spherical black-hole solutions of the Reissner-Nordstr\"{o}m-(anti-)de Sitter family, and the coordinate $f$ admits the range $f \in (0,\infty)$. Changing $f \rightarrow 1/r^2$ and rescaling time, we obtain
\begin{equation}\label{Inhomogeneous solution with Lambda better coordinates}
 \mathrm{d}s^2 = - \mathfrak{M}(r) \, \mathrm{d}t^2 + \frac{\mathrm{d}r^2}{\mathfrak{M}(r)} + r^2 \left( \mathrm{d}z^2 + \beta^2 \mathrm{d}\varphi^2 \right)
\end{equation}
with the same Maxwell field and
\begin{equation}\label{Inhomogeneous solution with Lambda transformed}
 \mathfrak{M}(r) = \frac{1}{r^2} + \frac{\Gamma}{r} - \frac{\Lambda}{3} r^2 \, ,
\end{equation}
where $\Gamma := \gamma/4$ and $[r]=m^{1/2}$. The metric coincides with (2.3) of \cite{Lemos+Zanchin} where the authors only discuss a negative cosmological constant but, in fact, both signs of $\Lambda$ are admissible. Let us explore its properties in more detail.

The spacetime has an axis at $r = 0$ where the circumferential radius of hoops $r,z,t=$ const. vanishes. The axis is conicity-free and at a finite proper distance from any point. The Kretschmann scalar
\begin{equation}\label{Kretschmann scalar}
  K = \frac{12 \Gamma^2}{r^6} + \frac{48 \Gamma}{r^7} + \frac{56}{r^8} + \frac{8}{3} \Lambda^2
\end{equation}
diverges at the axis which thus represents a singular source generating the magnetic field and splitting the admissible range of $r$ into positive and negative values---each range corresponds to an independent spacetime. However, $r$ always occurs squared in (\ref{Inhomogeneous solution with Lambda better coordinates}) apart from $\Gamma/r$ so that we can restrict ourselves to $r>0$ and both signs of $\Gamma$.

The solution is a warped product of a 2D black-hole spacetime (see, e.g., \cite{Lemos}) and $I\hspace{-0.13cm}R_2$ and it is type D apart from the hypersurface $r = -2/\Gamma$ (for a negative $\Gamma$) where it is type O. The magnetic field $F_{\mu\nu}F^{\mu\nu} = 2/r^4$ vanishes in the asymptotic region $r \rightarrow \infty$ where the solution approaches (anti-)de Sitter according to the sign of the cosmological constant.

The roots of the master function $\mathfrak{M}$ determine the positions of horizons---we convert $\mathfrak{M}$ to a single fraction and examine its numerator $3 + 3\Gamma r - \Lambda r^4$, which has a single extremum. For $\Lambda > 0$, there is thus always a single root corresponding to the cosmological horizon. For $\Lambda < 0$, there are either none or two or a single degenerate horizon if $\Gamma > \Gamma_{\mathrm{deg}}$, $\Gamma < \Gamma_{\mathrm{deg}}$, and $\Gamma= \Gamma_{\mathrm{deg}}$, respectively, with $\Gamma_{\mathrm{deg}} := - 4/3(-\Lambda)^{1/4}$. This critical value thus separates spacetimes with an inner and an outer black-string horizons from those with a naked singularity. It yields a special case where the master function has a single double root at $r=(-\Lambda)^{-1/4}$. It is of interest that this solution extends for the entire range $f = 1/r^2 \in (0,\infty)$ and it is thus different from the homogeneous solution (\ref{metric with deficit angle radial}), which only admits this particular value of $f = \sqrt{-\Lambda}$. The proper radial distance to the root is infinite while the circumferential radius is finite. The root corresponds to a degenerate horizon separating the asymptotic region $r \rightarrow \infty$ from the axis at $r=0$.

\begin{figure}
\centering
\includegraphics[width = 0.45\textwidth]{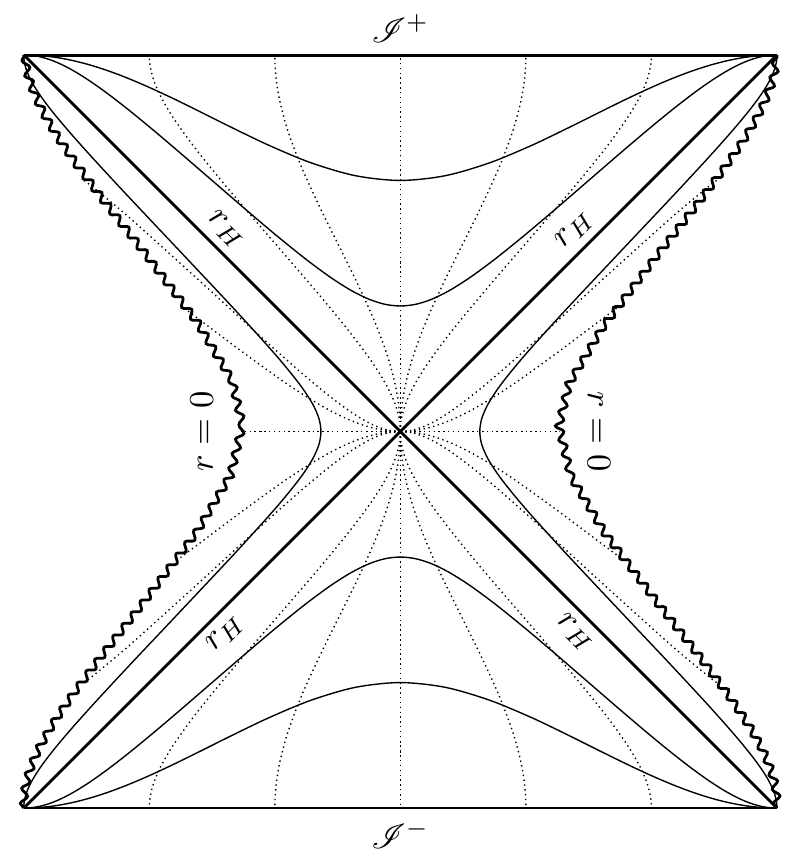}
\caption{A Penrose diagram for the metric (\ref{Inhomogeneous solution with Lambda better coordinates}) with $\Lambda>0$. There is always a single cosmological horizon located at $r_H$. Wavy, solid, and dotted lines represent singularities and constant $r$ and $t$, respectively.} \label{fig:grafyPhiPlus}
\end{figure}

\begin{figure*}[ht]
\centering
\subfloat[A naked singularity with $\Gamma > \Gamma_{\mathrm{deg}}$. \label{fig:grafyPhiKonf3D}]{\includegraphics[width = 0.25\textwidth]{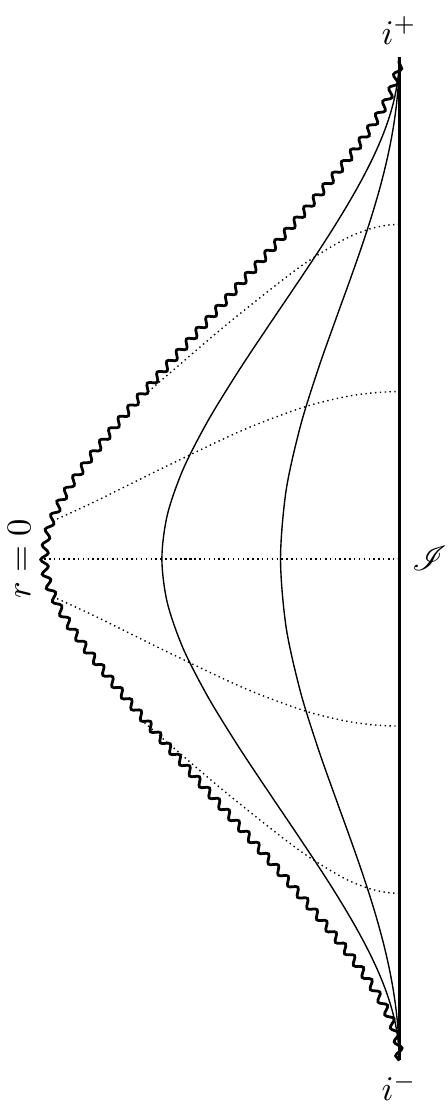}}
\hfill
\subfloat[The degenerate case featuring a double horizon at $r_H$, with $\Gamma = \Gamma_{\mathrm{deg}}$. To obtain a geodesically complete spacetime, we attach additional copies of this basic block at the top and bottom of the block along the horizon $r_H$. \label{fig:grafyPhiKonfContour}]{\includegraphics[width = 0.325\textwidth]{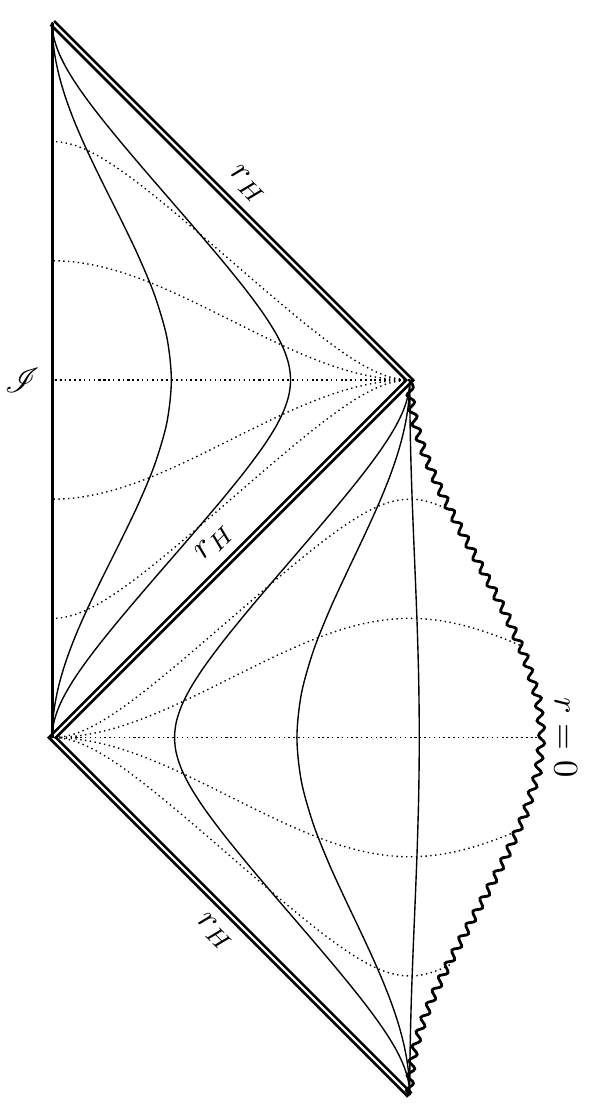}}
\hfill
\subfloat[The most general case with an inner and outer black-string horizons at $r_1$ and $r_2$, respectively, with $\Gamma < \Gamma_{\mathrm{deg}}$. Again, to get a full spacetime, we add copies of the basic block at the top and bottom along $r_2$. \label{fig:singsurfC}]{\includegraphics[width = 0.387\textwidth]{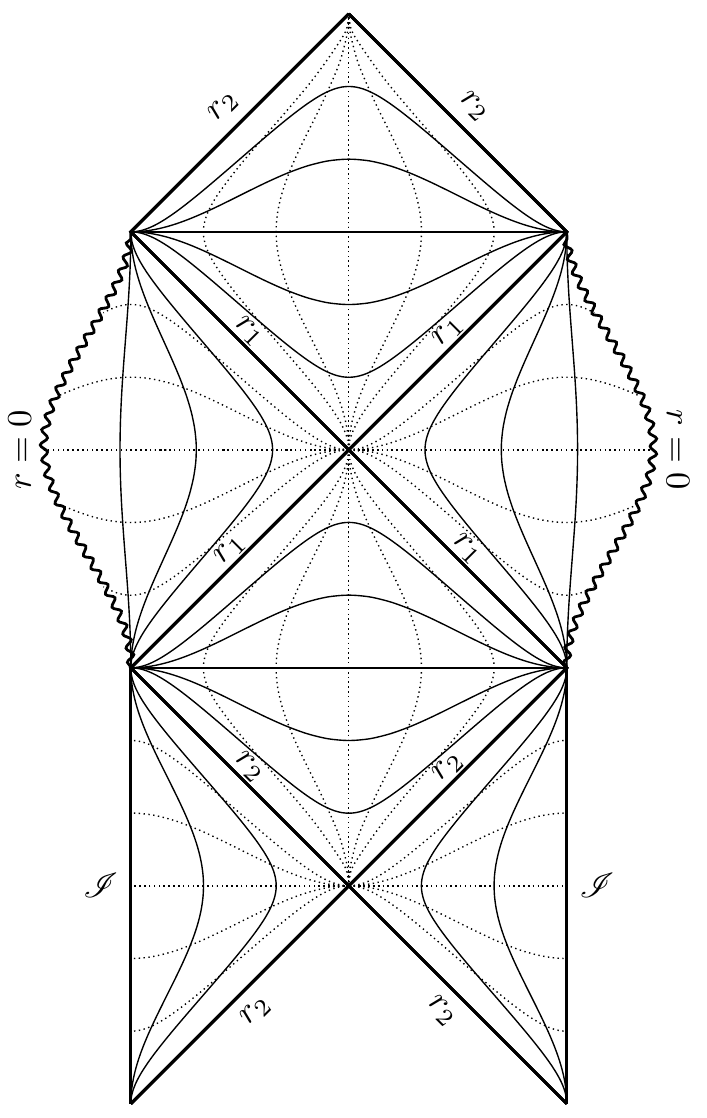}}
\caption{Penrose diagrams for the metric (\ref{Inhomogeneous solution with Lambda better coordinates}) with $\Lambda < 0$. Each point again represents a cylindrical surface and the wavy lines are singularities. Solid and dotted lines stand for constant $r$ and $t$, respectively. Depending on the parameters of the solution, we can have two horizons, one double horizon or no horizons at all.} \label{fig:grafyPhiMinus}
\end{figure*}

In Figures \ref{fig:grafyPhiPlus} and \ref{fig:grafyPhiMinus}, we present Penrose diagrams for all possible causal structures of the spacetime.
\section{“Inhomogeneous” symmetric solution with a vanishing \texorpdfstring{$\Lambda$}{L}}\label{Inhomogeneous_solution_without_Lambda}

\begin{figure*}
\centering
\subfloat[A single black-string horizon with $\Gamma < 0$ located at $r_H = -1/\Gamma$.]{\includegraphics[width = 0.325\textwidth]{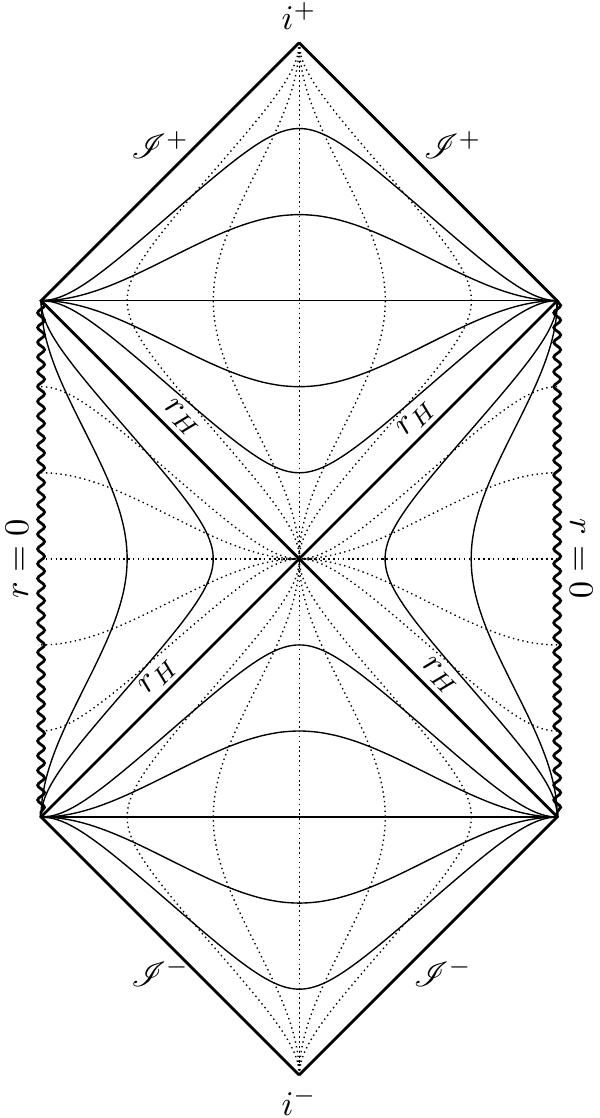}}\
\subfloat[A naked singularity with $\Gamma \geq 0$.]{\includegraphics[width = 0.325\textwidth]{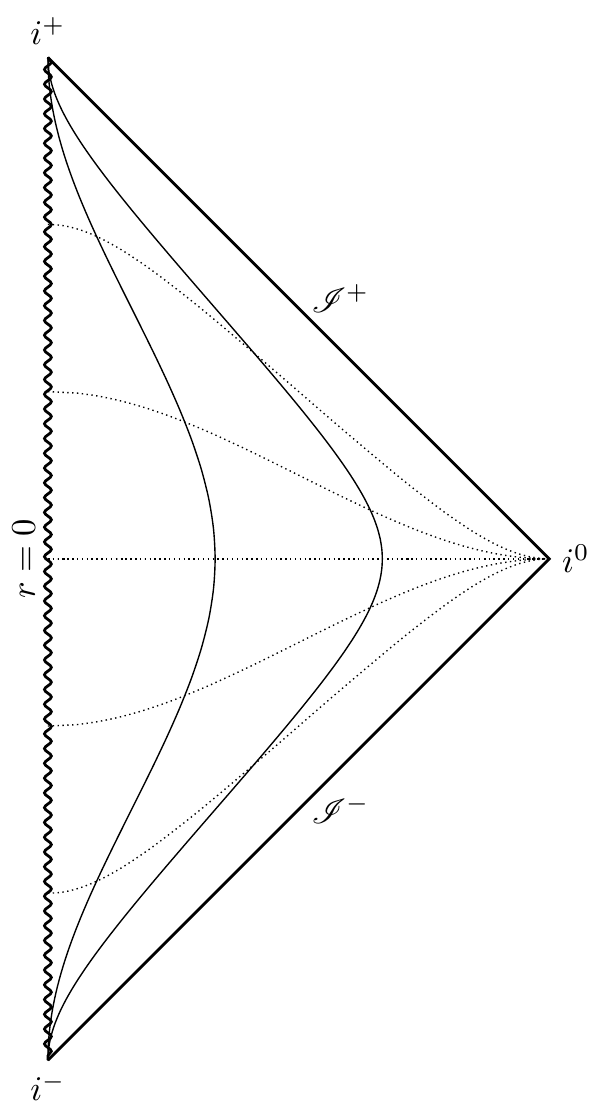}}
\caption{Penrose diagrams for the metric (\ref{Inhomogeneous solution with Lambda better coordinates}) with (\ref{master function with Lambda vanishing}), i.e., $\Lambda = 0$. Each point again represents a cylindrical surface, the wavy lines are singularities, solid lines represent constant $r$ and dotted lines constant $t$.} \label{fig:grafyPhi}
\end{figure*}

We now turn our attention to the limiting case of (\ref{Inhomogeneous solution with Lambda better coordinates}) when $\Lambda \rightarrow 0$, which we obtain simply by setting $\Lambda = 0$ in (\ref{Inhomogeneous solution with Lambda transformed}). The form of the metric does not change with the master function now simply reading
\begin{equation}\label{master function with Lambda vanishing}
 \mathfrak{M}(r) = \frac{1}{r^2} + \frac{\Gamma}{r}
\end{equation}
but the spacetime's properties have changed considerably. We now have $\mathfrak{M}(r) \rightarrow 0$ for $r \rightarrow \infty$ but the metric does not tend to the Minkowski flat metric since the Riemann tensor does not vanish. Both Maxwell and Kretschmann scalars vanish at radial infinity and they only diverge at $r=0$, which is thus still the location of a singular axis of a vanishing proper length. The possible horizon structure is also modified: we now have no horizons for $\Gamma \geq 0$ and the spacetime is static everywhere, and there is exactly one horizon for $\Gamma < 0$ located at $r = - 1/\Gamma$, see Figure \ref{fig:grafyPhi} for the corresponding Penrose diagrams.

The spacetime is again type D almost everywhere except for $r = -2/\Gamma$ (for a negative $\Gamma$) where it becomes type O, which further sets it apart from the solution (\ref{general line element original radial coordinate Wicked}) discussed above. However, it is not the singularity-free Bonnor-Melvin solution either since the axis is always singular here. There still are four Killing vector fields. The solution is a special case of (3.14) in \cite{Bronnikov+Santos+Wang} and of (27) in \cite{Richterek+Novotny+Horsky} (there is a typo in relation (31)).

The case of $\Lambda = \Gamma = 0$ is also interesting: the metric is conformastatic
\begin{equation}
 \mathrm{d}s^2 = - \frac{\mathrm{d}t^2}{r^2} + r^2 \left (\mathrm{d}r^2 + \mathrm{d}z^2 + \beta^2 \mathrm{d}\varphi^2 \right)
\end{equation}
and we still keep our magnetic field. There are no horizons and no new Killing vectors appear. The spacetime is type D everywhere. It is a special case of (3.15) in \cite{Bronnikov+Santos+Wang} with $q=1, b=0$.
\section{Final considerations and conclusions}
Let us briefly visit the most general case of $B \not = C$ regarding the system  (\ref{Einstein t radial})-(\ref{Einstein r radial}). It turns out that the equations still can be separated similarly to the axial case to produce a single third-order equation for $f$
\begin{eqnarray}
     &f'''f'& - 2(f'')^2 + f''\left(6f(\Lambda+f^2) + \frac{(f')^2}{f}\right) \nonumber\\
     && - (f')^2(11f^2 + 9\Lambda) - 4f^2(\Lambda+f^2)^2=0. \label{full equation for f radial}
\end{eqnarray}
After solving this equation we use $f$ to find the metric functions from the following expressions
\begin{eqnarray}
  A' &=& -2 \frac{f''}{f'} + 4 \frac{f'}{f} + 4\frac{f}{f'}(\Lambda+f^2),\label{differential equation for A radial}\\
  B' &=& -\frac{f'}{f} \pm \sqrt{4\frac{f''}{f}-7\left(\frac{f'}{f}\right)^2-4(\Lambda+f^2)},\label{differential equation for B radial}\\
  C' &=& -\frac{f'}{f} \mp \sqrt{4\frac{f''}{f}-7\left(\frac{f'}{f}\right)^2-4(\Lambda+f^2)}. \label{differential equation for C radial}
\end{eqnarray}
If the square root in (\ref{differential equation for B radial}) and (\ref{differential equation for C radial}) vanishes, we obtain the previous, symmetric solution. Since (\ref{full equation for f radial}) does not involve the independent variable, we can reduce the order of the equation by one similarly to (\ref{full equation for w}) to obtain
\begin{eqnarray}
    &w \ddot{w}& - \dot{w}^2 + \left[\frac{w}{f} + 6 f \left(f^2 + \Lambda\right)\right] \dot{w} \nonumber\\
    &&- 8 \left(f^2 + \Lambda\right)^2 f^2 - 2 w \left( 11 f^2 + 9 \Lambda \right) = 0,\label{full equation for w radial}
\end{eqnarray}
with dots denoting derivatives with respect to $f$ now and $w(f) := [\mathrm{d}f(r)/\mathrm{d}r]^2 $. Unfortunately, we have so far not been able to solve the above equations analytically. They seem to require a numerical approach, which thus goes beyond the methods we preferred in the present text. We stress that the spacetimes resulting as solutions of both the above equation and (\ref{full equation for w}) are not included in \cite{Bronnikov+Santos+Wang} since its authors consider either the cosmological constant, or an electromagnetic field but never the two together.

In this paper we studied a system of Einstein-Maxwell equations describing the gravitational and magnetic fields of a cylindrically symmetric static system where the magnetic field is perpendicular to the axis of symmetry. We explored in detail the properties of several classes of solutions both with and without a cosmological constant and pointed out a possible way forward by reducing the set of Einstein-Maxwell equations to a single, second-order non-linear differential equation.
\begin{acknowledgments}
J.V. was supported by Charles University, project GAUK 80918. M.Z. acknowledges funding by GACR 21-11268S.
\end{acknowledgments}


\begin{thebibliography}{99}
\bibitem{Bonnor} W. B. Bonnor, Proc. Phys. Soc., London, Sect. A \textbf{67}, 225 (1954).
\bibitem{Melvin} M. A. Melvin, Phys. Lett. \textbf{8}, 65 (1964).

\bibitem{Yoo+Harada+Okawa} C-M. Yoo, T. Harada and H. Okawa, Class. Quantum Grav. \textbf{34}, 105010 (2017).
\bibitem{East} W. E. East, Phys. Rev. Lett. \textbf{122}, 231103 (2019).
\bibitem{Copeland+Kibble} E. J. Copeland and T. W. B. Kibble, Proc. R. Soc. A \textbf{466}, 623 (2010).
\bibitem{Thorne} K. S. Thorne, in \emph{Magic without Magic:  John Archibald Wheeler}, edited by J. Klauder (Freeman, San Francisco, 1972).
\bibitem{Hod} S. Hod, Eur. Phys. J. C \textbf{80}, 982 (2020).
\bibitem{Zofka} M. \v{Z}ofka, Phys. Rev. D \textbf{99}, 044058 (2019).
\bibitem{Astorino} M. Astorino, J. High Energy Phys. \textbf{86}, 2012 (2012).
\bibitem{Lim} Y. K. Lim, Phys. Rev. D \textbf{98}, 084022 (2018).
\bibitem{Vesely+Zofka} J. Vesel\'{y} and M. \v{Z}ofka, Phys. Rev. D \textbf{100}, 044059 (2019).
\bibitem{Bronnikov+Santos+Wang} K. Bronnikov, N. Santos and A. Wang, Class. Quantum Grav. \textbf{37} 113002 (2020).
\bibitem{Vesely+Zofka_geodesics} J. Vesel\'{y} and M. \v{Z}ofka, Gen. Relativ. Gravit. \textbf{51}, 156 (2019).
\bibitem{Plebanski+Hacyan} J. F. Pleba\'{n}ski and S. Hacyan, J. Math. Phys. \textbf{20}, 1004 (1979).
\bibitem{Griffiths+Podolsky} J. B. Griffiths and J. Podolsk\'{y}, \emph{Exact Space-Times in Einstein's General Relativity} (Cambridge University Press, 2009).
\bibitem{Lemos+Zanchin} J. P. S. Lemos and V. T. Zanchin, Phys. Rev. D \textbf{54}, 3840 (1996).
\bibitem{Lemos} J. P. S. Lemos, Class. Quantum Grav. \textbf{12} 1081 (1995).
\bibitem{Richterek+Novotny+Horsky} L. Richterek, J. Novotn\'{y} and J. Horsk\'{y}, Czech. J. Phys. \textbf{50}, 925 (2000).
\end{thebibliography}
\end{document}